\documentclass[article]{JHEP3}

\usepackage{amsmath,epsfig}
\usepackage[latin1]{inputenc}
\usepackage{amssymb,amsfonts}

\title{Schwinger pair production in $AdS_2$}

\preprint{\hepth{0501169}\\LPTHE-05-02\\LPTENS-05-04}

\author{ B.~Pioline$^{\clubsuit\spadesuit}$ and J.~Troost$^\spadesuit$\\ \\
$\clubsuit$~LPTHE, Unit\'e mixte du CNRS 
et des Universit\'es Paris 6 et 7, bo\^\i te 126,\\
4 place Jussieu,  75252 Paris cedex 05, France
\\ \\
$\spadesuit$~LPTENS, Unit\'e mixte du CNRS et de l'Ecole
Normale Sup\'erieure, D\'epartement de Physique de l'ENS, 24 rue Lhomond,
75231 Paris cedex 05, France
\\ \\
{\tt E-mail:
pioline@lpthe.jussieu.fr, troost@lpt.ens.fr}
}

\abstract{
We analyze the pair production of charged particles in two-dimensional 
Anti-de Sitter space ($AdS_2$) with a constant, uniform
electric field. We compute the production rate both 
at a semi-classical level, viewing Schwinger pair production 
as a tunneling event, and at the full quantum level, by extracting
the imaginary part of the one-loop amplitude. In contrast to the usual
Schwinger pair production in flat space, 
pair production in $AdS_2$ requires a sufficiently large
electric field $E^2> M^2+1/4$ in order to 
overcome the confining effect of the $AdS$ geometry --
put in another way, the presence of
an electric field $E$ raises the Breitenlohner-Freedman bound to
$M^2 > -1/4 + E^2$. For $E$ greater than this threshold, the vacuum
is unstable to production of charged pairs in the bulk. 
We expect our results to be helpful in constructing supersymmetric 
$AdS_2\times S^2$ perturbative string vacua,
which enter in the near-horizon limit of extremal charged black holes.
Although the generalized  Breitenlohner-Freedman bound is obeyed in 
these cases, production of BPS particles at threshold is possible
and relevant for $AdS_2$ fragmentation.}

\makeatletter
\renewcommand{\subsubsection}{\@startsection{subsubsection}{3}{0mm}{-\baselineskip}{0.5\baselineskip}{\normalfont\normalsize\it}}
\makeatother

\newcommand{\pa}{\partial}
\newcommand{\nn}{\nonumber}
\newcommand{\eps}{\epsilon}
\newcommand{\Real}{\mathbb{R}}
\newcommand{\Zint}{\mathbb{Z}}

\newcommand{\Tr}{\mbox{Tr}}
\newcommand{\Li}{\mbox{Li}}

\def\bea{\begin{eqnarray}}
\def\eea{\end{eqnarray}}
\def\be{\begin{equation}}
\def\ee{\end{equation}}
\def\ba{\begin{align}}
\def\ea{\end{align}}
\def\bse{\begin{subequations}}
\def\ese{\end{subequations}}
\def\1F1{{}_1\!F_1}
\def\2F0{{}_2\!F_0}

\begin{document}

%\maketitle 
%\setcounter{tocdepth}{2}
%\tableofcontents

\section{Introduction}

Two-dimensional anti-de Sitter space $AdS_2$ plays a central r\^ole in
the physics of extremal four-dimensional Reissner-Nordstr\"om black holes.
In Einstein-Maxwell gravity, the near-horizon geometry of an extremal dyonic
black hole reduces to the Bertotti-Robinson space-time 
$AdS_2\times S^2$ \cite{Gibbons:1982ih}, each factor supporting an
electric (resp. magnetic) flux proportional to the electric (resp. magnetic)
charge of the black hole. In type IIA (IIB) string theory, the same
space-time, tensored with a a Calabi-Yau three-fold $Y$ of fixed complex
(K\"ahler) structure, appears as the universal near-horizon geometry for all
``three-charge'' supersymmetric black holes \cite{Ferrara:1995ih,
Moore:2004fg}. In the non-extremal case, the near-horizon geometry is 
still locally $AdS_2\times S^2$, although the global structure 
differs \cite{Maldacena:1998uz,Spradlin:1999bn,Kim:1998wy}.

Since the macroscopic entropy of the black hole is
proportional to the area of the sphere $S^2$, it is natural to expect 
that a better understanding of string theory in $AdS_2\times S^2$ may
shed light on the microscopic degrees of freedom of extremal black holes.
Indeed, the superconformal quantum mechanics of charged D0-branes on  
$AdS_2\times S^2$ has been recently shown to reproduce
the entropy of a class of extremal black holes \cite{Gaiotto:2004pc}.
Unfortunately, the study of string theory in the relevant backgrounds
is hampered by the existence of Ramond fluxes (see however 
\cite{Berkovits:1999zq} for recent progress).

It is therefore interesting to note that $AdS_2\times S^2$
admits an exact conformal field theory description as an asymmetric coset
$[Sl(2) \times SU(2)]/U(1)\times U(1)$ \cite{Giddings:1993wn,Johnson:1994kv}.
The latter can be embedded both in the heterotic string, where the
electromagnetic flux is provided by the 10-dimensional gauge fields,
or in the type II string, so long as the fluxes originate from the
Neveu-Schwarz sector. In particular, the analogue of charged 
D0-brane probes now are perturbative string states, which can be analyzed by
ordinary field theoretic methods.
Unfortunately, this conformal field theory remains little understood,
as it involves a coset by an hyperbolic generator (see \cite{Israel:2004vv} 
for a recent discussion).

%More recently, this has been found to hold for
%black holes with vanishing classical entropy as well, upon including
%a class of higher derivative interactions
%\cite{Dabholkar:2004yr,LopesCardoso:1998wt,Sen:1997is,Dabholkar:2004dq}. 

%Understanding string theory in $AdS_2\times S^2$ backgrounds may thus
%be a key step in uncovering the microscopic origin of the Hawking entropy.

In addition to these technical difficulties, quantum gravity on $AdS_2$ 
raises more general puzzles, which have thwarted a direct application of the
holographic ideas \cite{Strominger:1998yg,Gibbons:1998fa} (see \cite{Thompson:2003fz,
Strominger:2003tm,Verlinde:2004gt} for recent progress).
In contrast to higher-dimensional
anti-de Sitter space, global $AdS_2$ has two disconnected time-like 
boundaries, much like the worldsheet of an ordinary open string.
Indeed, global $AdS_2$  can be viewed as the $Sl(2,\Real)$ invariant 
ground state of two-dimensional (Liouville) gravity on the
strip \cite{D'Hoker:1983is}. Just like  perturbative open strings
are liable to splitting and joining,
$AdS_2$ can fragment, or rather nucleate
baby universes \cite{Maldacena:1998uz}. Such
branching geometries can be understood as the near-horizon limit of a 
multi-centered configuration of supersymmetric RN black holes
and form a large moduli space of 
zero-energy configurations \cite{Britto-Pacumio:1999ax}.
The fragmentation process can be described by an Euclidean wormhole
configuration
describing the semi-classical tunneling from one $AdS_2$ throat of
charge $Q_0>0$ to two throats of charges $(Q_1>0,Q_2=Q_0-Q_1>0)$
\cite{Brill:1991rw}. Interestingly, the production rate turns out to be
proportional to the difference of the Hawking entropies 
of the initial and final state.

When the charge of one black hole is much smaller compared to the
others, the process of fragmentation can be understood as the
Schwinger pair creation of D0-branes, charged under the 
electric flux threading $AdS_2$.
The semi-classical computation has been outlined in the 
BPS case in \cite{Maldacena:1998uz}, producing agreement with 
the large charge case above. In this work, we give a careful
analysis of the one-loop vacuum amplitude for charged particles
in $AdS_2$ in the presence of a constant, uniform electric 
field. Our computation is closely related, but differs by 
the type of analytic continuation used, to the computation of
the one-loop amplitude for a charged particle in the planar patch
of $dS_2$ performed long ago in \cite{Comtet:1984mm}.
As usual, the real part of the one-loop amplitude carries
information about dispersive effects, while the imaginary
part gives the (tree-level) production rate for charged particles
in the electric field. The latter is expected to vanish in 
supersymmetric backgrounds, except for BPS states which may be produced
abundantly and can have a large gravitational backreaction, leading
to the possibility of baby universe formation.

The outline of this paper is as follows. In Section 2 we discuss
some aspects of propagation of a charged particle in $AdS_2$,
give a semi-classical picture of Schwinger pair creation,
as a tunneling process in the potential governing the radial 
motion, and perform a ``mini-superspace'' computation of 
the one-loop amplitude. In Section 3,  we compute the one-loop
vacuum amplitude by analytic continuation from the Landau 
problem on the hyperbolic plane $H_2$. Section 4 contains
a Summary and Discussion. An alternative regularization
scheme is outlined in Appendix A, and gravitational
corrections are discussed in Appendix B.

\section{Semi-classics of a charged particle in $AdS_2$}
In this section, we discuss some semi-classical aspects of the propagation
of a charged particle in $AdS_2$, in the presence of a constant electric 
field. In Poincar\'e coordinates, the space-time metric reads
\be
ds^2 = a^2 \frac{-dt^2 + dy^2}{y^2}
\ee
where the dimensionful parameter $a$ sets the scale of the curvature,
$R=-2/a^2$ (check sign), and the coordinate $y$ is restricted to 
positive values only. While the locus $y=0$ corresponds to the 
boundary of $AdS^2$, the geometry can be continued across 
light-like infinity in the $(t,y)$ plane by defining ``strip''
coordinates
\be
\tau=\arctan(y+t)-\arctan(y-t)\ , \quad
\sigma=\arctan(y+t)+\arctan(y-t)
\ee
so that the strip $0\leq \sigma\leq \pi, \tau\in\Real$ now cover the  
global $AdS_2$ geometry, with metric
\be
ds^2 = a^2 \frac{-d\tau^2 + d\sigma^2}{\sin^2\sigma}
\ee
In particular, the locus $y=0$ describes the
time-like boundary at $\sigma=0$ for a finite global time
$-\pi\leq \tau\leq \pi$, while $y\to\infty$ barely touches the other
boundary at $\sigma=\pi$  at global time $\tau=0$. 
The global geometry of $AdS_2$ can be covered by two infinite family
of Poincar\'e patches. 

\subsection{Classical trajectories of a charged particle in $AdS_2$}

Let us now introduce a constant electric field $F=E \omega /a^2$ 
proportional to the volume element, $\omega = a^2 dt dy/y^2 =
a^2 d\tau d\sigma / \sin^2\sigma$. It is convenient to choose different
gauges in the Poincar\'e and global coordinates,
\be
A = E\ \frac{dt}{y}\ ,\quad 
\tilde A = E \frac{d\tau}{\tan\sigma}
\ee
which differ by a gauge transformation,
\be
A-\tilde A= E d\lambda, \ ,\quad
\lambda = \frac12 \log \frac{1+(t+y)^2}{1+(t-y)^2} = 
\log \frac{ \cos[(\tau-\sigma)/2]} { \cos[(\tau+\sigma)/2]} 
\ee 
A charged particle propagating in the Poincar\'e patch of 
$AdS_2$ in the presence of an electric field is thus described by the action
\be
\label{acp}
S[y,t] = \int ds \ \left\{ \frac{a^2}{2 \rho y^2} \left[ 
- \left( \frac{dt}{ds} \right)^2 + \left( \frac{dy}{ds} \right)^2 \right]
+ \frac{E}{y}\  \frac{dt}{ds} \ -\ \rho\ \frac{M^2}{2} \right\} 
\ee
where the Lagrange multiplier $\rho$ enforces the mass-shell condition.
Setting $\rho=1$, the world-line Hamiltonian becomes
\be
\label{hpoin}
H_{\rm Poinc}= \frac{y^2}{a^2} 
\left[ p_y^2 - \left( p_t - \frac{E}{y} \right)^2 \right] + M^2 
\equiv 0
\ee
and is required to vanish due to reparameterization invariance.
In this expression, the momenta canonically conjugate to $y,t$ are given by
\be
p_y = \frac{a^2}{y^2} \frac{dy}{ds}\ ,\quad
p_t = - \frac{a^2}{y^2} \frac{dt}{ds} + \frac{E}{y} \ .
\ee
Since $p_t$ is a conserved quantity (the Poincar\'e energy), the motion in 
the radial coordinate $y$ is simply governed by a one-dimensional potential
$V(y)$,
\be
\label{hpoinc}
p_y^2 + V(y) = 0\ ,\quad 
V(y) =  \frac{M^2}{a^2\ y^2}- \left( p_t - \frac{E}{y} \right)^2
\ee
From now on, we will set the dimensionful parameter $a=1$.
From this expression, it is easy to infer the following facts (see Figure 1):
\begin{itemize}
\item[i)] For small electric field $E^2<M^2$, the charged particle (electron,
for short)
stays at a finite distance $y\geq (E+ \eps M)/p_t$ from the boundary, where 
$\eps=\pm 1$ denotes the sign of $p_t$, and reaches the horizon of the
Poincar\'e patch at $y=\infty$.
\item[ii)] For large electric field $E^2>M^2$ and $E p_t>0$, the motion
consists of two branches: one staying at a finite distance 
$y\geq (E+\eps M)/p_t$ from the boundary and reaching the horizon, 
corresponding to the motion of the electron, and the other,
confined near the boundary at $y<(E-\eps M)/p_t$, corresponding
to the motion of the positron, being emitted and reabsorbed by the
boundary of $AdS_2$. As we shall explain in Section 2.3,
tunneling between these two branches corresponds to Schwinger pair production
in the bulk of $AdS_2$.
\item[iii)] For large electric field $E^2>M^2$ and $E p_t<0$, the motion
covers the entire half axis $y>0$, and consists of two branches 
extending from the boundary to the bulk, and conversely.
These trajectories can be viewed as emission/absorption of 
an electron from the boundary of $AdS_2$.
\item[iv)] For a critical electric field $E^2=M^2$, the positron
trajectory in case ii) disappears into the boundary, while the electron
trajectory in case iii) is tangent to the boundary. This case is 
relevant for BPS particles in a supersymmetric $AdS_2$ background.
\end{itemize}

\FIGURE{
\epsfig{file=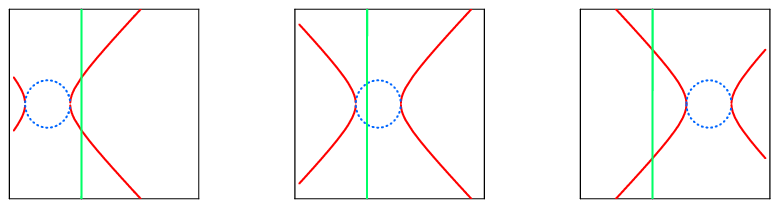,height=3.5cm} 
\caption{Classical trajectories of a charged particle in the 
Poincaré patch of $AdS_2$. The latter lies to the right of the
green line $y=0$. The solid red (resp. dotted blue)
line shows trajectories in real (resp. imaginary) proper time.
The right (resp. left) branch corresponds to the electron (resp.
positron) trajectory.
For weak electric field $E^2<M^2$ ({\it center}), the vacuum is stable.
For strong electric field $E^2>M^2$, particles with $E p_t>0$
can be pair produced in the bulk ({\it right}), unlike those with 
$E p_t<0$ ({\it left}).}}

Similarly, in global coordinates, the motion along the spatial
direction $\sigma$ is governed by the potential
\be
\label{hglob}
H_{\rm glob} = {\sin^2 \sigma} \left[ p_\sigma^2 + V(\sigma) \right] \equiv
 0 \ ,\quad 
V(\sigma) =
\frac{M^2}{\sin^2 \sigma} - \left(p_\tau - \frac{E}{\tan \sigma}\right)^2
\ee
where the canonical momenta are given by
\be
p_\sigma = \frac{a^2}{\sin^2 \sigma} \frac{d\sigma}{ds}\ ,\quad
p_\tau  = - \frac{a^2}{\sin^2 \sigma} \frac{d\tau}{ds} + \frac{E}{\tan\sigma}
\ee
It is now easy to determine the semi-classical spectrum  (see Figure 2)
for a fixed value of the global energy $p_\tau$:
\begin{itemize}
\item[i)]
For small electric field $E^2<M^2$, the electron (or the positron)
is confined into the bulk of $AdS_2$. The semi-classical spectrum 
can be obtained by the Bohr-Sommerfeld quantization rule (generalizing
the analysis for zero electric field in \cite{Nakatsu:1998st}), and
consists of discrete states of energy
\be
| p_\tau| = \sqrt{M^2- E^2} + n + \frac12
\ee
where $n$ is a non-negative integer. 
\item[ii)] When the electric field becomes 
greater than the critical value $E^2>M^2$, the potential barriers
at $\sigma=0,\pi$ which prevented the particle to reach the boundaries
become potential wells, while a potential barrier appears at finite
value of $\sigma$. The particle is now
confined to either of the two boundaries, and quantum tunneling 
under the barrier can be interpreted as Schwinger pair creation
in the bulk.
\end{itemize}

\FIGURE{
\epsfig{file=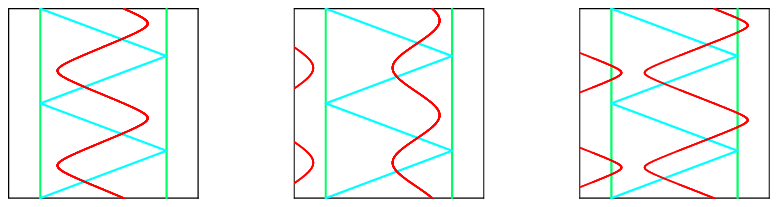,height=3.5cm} 
\caption{Classical trajectories of a charged particle in global $AdS_2$.
Green lines denote the boundaries $\sigma=0,\pi$, blue lines the
horizon of the Poincaré patch, and red lines show
trajectories in real proper time.
The electric field is increased from $E=0$ (on the left) to
$E^2>M^2+\frac14$ (on the right). 
}}

\subsection{Spectrum generating $Sl(2,\Real)$ symmetry}

In order to gain further insight on these trajectories, it is useful
to note that the Hamiltonians \eqref{hpoinc},\eqref{hglob} exhibit an 
$Sl(2,\Real)$ symmetry, as a consequence 
of the isometries of $AdS_2$. In Poincaré
coordinates, the canonical generators are
\bea
\label{sl2poinc}
L_0 &=& t \ p_t  + y \ p_y \\
L_+ &=& i \ p_t \\
L_- &=& i \left[ \ (t^2+y^2)\ p_t + 2\ ty \ p_y - 2 E y \right]
\eea
while, in global coordinates,
\bea
\label{sl2glob}
L_0 &=& - i ( \cos\tau\sin\sigma \pa_\sigma  + 
\cos\sigma \sin\tau \pa_\tau) + E \sin\sigma\sin\tau\\
L_+ &=&  -\sin\sigma\sin\tau \pa_\sigma +
(1+\cos\sigma\cos\tau)\pa_\tau 
+i E \sin\sigma \cos\tau\\
L_- &=& \sin\sigma \sin\tau \pa_\sigma  +
(1-\cos\sigma\cos\tau)\pa_\tau  - i E \sin\sigma \cos\tau
\eea
In either case, the Hamiltonian is related to the quadratic
Casimir\footnote{Although Eqs \eqref{sl2poinc},\eqref{sl2glob} are
classical, their quantum counterpart is simply obtained by replacing
the momenta $p_i$ by $-i \pa_i$, preserving the ordering shown; as
a result, the resulting quadratic Casimir is shifted to $C=M^2-E^2 +\frac14$,
and $M^2$ is shifted to $M^2+\frac14$.}
of the $Sl(2)$ representation via
\be
C = -L_0^2 - \frac12 \left( L_+ L_- + L_- L_+ \right) = M^2 - E^2
\ee
Notice in particular that the Poincaré energy $p_t=-i L_+$ and global
energy $p_\tau=-i (L_+ + L_-)/2$ correspond to parabolic and elliptic
generators, respectively. From our semi-classical discussion in the
previous section, we find that the spectrum consists of  
discrete representations of $Sl(2,\Real)$
for states below the pair production
threshold $E^2<M^2$, and continuous
representations for states above.

Eliminating the momenta $p_t, p_y$ from the three
equations \eqref{sl2poinc}, one readily obtains the equation for the
intrinsic trajectory of a charged particle in Poincaré coordinates, 
\be
\left( y + \frac{E}{i L_+} \right)^2 - 
\left(t - \frac{L_0}{L_+} \right)^2 = \frac{M^2}{L^2_+}
\ee
Thus, charged particles follow a branch of hyperbola in the Poincar\'e
patch, centered at a radial distance $y=-E/p_t$. This generalizes
to the Lorenzian, charged case the well-known fact that (neutral) geodesics
on the hyperbolic plane are circles centered on the boundary at $y=0$.
Similarly, the intrinsic trajectory in global coordinates can be
obtained by eliminating $p_\tau,p_\sigma$ from  \eqref{sl2glob}:
\be
i(L_+-L_-)\cos\tau +2 L_0 \sin \tau = 
2 E \sin \sigma - i (L_++L_-) \cos\sigma 
\ee
Notice in particular that all trajectories have the same period $2\pi$
in global time $\tau$.

\subsection{Tunneling and semi-classical Schwinger pair production}
\label{smsch}
As first discussed in \cite{Brezin:1970xf}, and reviewed e.g. in
\cite{Pioline:2003bs}, Schwinger pair production can be understood 
semi-classically as a tunneling process between two classically
allowed regions of the potential describing the relativistic motion of 
the electron in the presence of an electric field. For example, in
two-dimensional Minkowski space with a constant electric field $E$,
the motion along the spatial coordinate $x$, at fixed energy $p_t$,
is given by the inverted harmonic potential
\be
p_x^2 + V(x) = 0\ ,\quad
V(x) = M^2 - (p_t - E x)^2
\ee
Particles to the right of the potential barrier at $x=p_t/E$ 
correspond to electrons, while particles to the left are positrons.
Tunneling under the barrier corresponds to a stimulated emission
of positron-electron pairs. The tunneling rate can be computed
semi-classically by evolving the wave function in imaginary 
proper time, which amounts to flipping the sign of the potential.
There is now an Euclidean trajectory relating the two
turning points, with classical action
\be
S_{cl} = \int_{(p_t-M)/E}^{(p_t+M)/E} \sqrt{V(x)} \ dx = \frac{\pi M^2}{2E}
\ee
The semi-classical pair creation rate, either for stimulated or spontaneous
processes, is therefore given by 
$\exp(-2 S_{Cl})$, in agreement with Schwinger's classic result.
Notice that the Euclidean instanton can be viewed as the motion
of a charged particle in a constant {\it magnetic} field during
a Larmor period.

By the same token, the pair production rate in $AdS_2$ may be 
computed semi-classically by evaluating the action of the instanton
which controls the tunneling between the electron and positron
branches, in the case (ii) above. As in flat space, 
the propagation of a charged particle on $AdS_2$ in imaginary proper 
time amounts to a real-time evolution under the Hamiltonian
\be
\label{hpoinb}
H_{B}= \frac{y^2}{a^2} 
\left[ p_y^2 + \left( p_x - \frac{E}{y} \right)^2 \right] 
\ee
upon identifying $B=E, H_B \equiv M^2$. Formula \eqref{hpoinb} is now
the Hamiltonian of a charged particle
on the Euclidean hyperbolic plane $H_2$, with metric and
magnetic field 
\be
ds^2 = a^2 \ \frac{dx^2 + dy^2}{y^2}\ ,\quad A= B \frac{dx}{y}
\ee 
The Landau problem on $H_2$ was discussed in detail in
\cite{Comtet:1986ki} (see also e.g. 
\cite{Grosche:1989kj,Antoine:1990ew}), 
with the following semi-classical results:
\begin{itemize}
\item[a)] Classical trajectories are circles in the $(x,y)$ plane,
centered at $y=B/p_x$. For small magnetic field $B^2<H$, they
are open trajectories which intersect the boundary $y=0$; for 
large magnetic field $B^2>H\geq 0$, they are closed 
trajectories in the bulk of $H_2$.
\item[b)] Open trajectories reach the boundary in infinite proper time,
  and have an infinite action. They correspond to a continuum of
  states with energy $H\geq B^2$.
\item[c)] Closed trajectories have a finite action
\be
\label{sc1}
S = \pi \left( B - \sqrt{B^2-H}  \right)
\ee
By the Bohr-Sommerfeld condition, they correspond to a finite number 
of discrete states with energy\footnote{It is perhaps worth noting
that, in contrast to the flat space case, the area enclosed by 
each trajectory is not a multiple of a fixed quantum, but rather
$A_n = \pi \left( n+ \frac12 \right) \slash
\left[B - \frac12\left( n+\frac12 \right) \right]$.}
\be
H_n =  \left( n + \frac12 \right) \left( 2B- n - \frac12 \right)
\ee
with $0<n+1/2<B$.
\end{itemize}
Cases b) and c) are the analytic continuation of the
electric trajectories i) and ii). We conclude that 
Schwinger pair production in the bulk of $AdS_2$, corresponding
to the tunneling between the two electric trajectories of type ii), 
is mediated by an instanton trajectory of type c), with rate
\be
\label{gbulk}
\Gamma_{bulk} = \exp\left[ -2\pi \left( E - \sqrt{E^2 - M^2} \right) \right]
\ee 
As we shall see, this semi-classical result
agrees with the quantum analysis in Section 3.
It is also easy to check that \eqref{gbulk} is in agreement with the 
result of \cite{Maldacena:1998uz} for the action of a $(D-2)$
spherical brane in $AdS_D$, with $D=2$.

Finally, let us compare the pair production rates in 
global $AdS_2$ and in the Poincar\'e patch. The WKB
actions controlling the tunneling process are given by
\be
S_{\rm{Poinc.}} = \oint \ p_y dy\ ,\quad
S_{\rm{Glob.}} =  \oint p_\sigma d\sigma
\ee
where the integral runs over a closed periodic trajectory\footnote{
The standard WKB prescription of integrating over {\it half} a period
would not be gauge-invariant.} 
in imaginary time, and $p_y = \sqrt{V(y)}$ 
(resp. $p_\sigma = \sqrt{V(\sigma)})$ is the momentum derived from
the potential $V(y)$ in \eqref{hpoinc} (resp. $V(\sigma)$ in \eqref{hglob})
respectively, and $p_y = \sqrt{V(y)}$ (resp. $p_\sigma = \sqrt{V(\sigma)})$.
Noting that the Lagrangians in Poincar\'e and global coordinates differ
by a total derivative, we have
\be
\label{eqs}
p_y dy + p_t dt - H_{\rm{Poinc.}} ds 
= p_\sigma d\sigma + p_\tau d\tau - H_{\rm{Glob.}} ds + E d\lambda
\ee
The Hamiltonians $H_{\rm{Poinc.}}$ and $H_{\rm{Glob.}}$ are equal
off-shell, hence cancel from this equality. Let us now integrate
\eqref{eqs} along a closed Euclidean trajectory: since $p_t$ and $p_\tau$ are
conserved quantities and their conjugate variables are periodic, we find
that only the first term on either side remains, and conclude that
the semi-classical tunneling rates are the same in the Poincar\'e
patch and in global $AdS_2$. This suggests that the equivalence of
the Poincar\'e and global vacuum demonstrated for neutral particles
in \cite{Danielsson:1998wt} carries over to the charged case.

\subsection{Periodic trajectories and mini-superspace path integral}
Quantum mechanically, the pair creation rate can be obtained as 
the imaginary part of the
one-loop vacuum amplitude of a charged particle in $AdS_2$,
with action \eqref{acp}. Before turning to a quantum field theory
computation in Section 3, it is instructive 
to study a ``mini-superspace'' version
of this computation, following the approach initiated in 
\cite{Affleck:1981bm} for monopole pair creation in a magnetic field
(see \cite{Berkooz:2004re} for a review of this technique).

Recall that the logarithm of the one-loop vacuum amplitude
(or free energy, for brevity) is given by 
the Euclidean path integral
\be
W_B = - \int_{0}^\infty \frac{d\tilde\rho}{\tilde\rho} 
\int [dx(s)]\  [dy(s)] 
\ \exp( - S_E )
\ee
where $S_E$ is the Euclidean action
\be
\label{acpb}
S[x,y,\rho] = \int ds \ \left\{ \frac{a^2}{2 \rho y^2} \left[ 
+\left( \frac{dx}{ds} \right)^2 + \left( \frac{dy}{ds} \right)^2 \right]
+ \frac{B}{y}\  \frac{dx}{ds} \ -\ \rho\ \frac{M^2}{2} \right\} 
\ee
and the path integral is restricted to configurations periodic
in imaginary time $s$ with period $2\pi$. For later convenience, we
shall use the gauge $\rho(s)=\tilde \rho / y(s)$. 
The idea now 
is to truncate the path integral to a family of closed
off-shell configurations which contain the instantons mediating the
pair creation process, namely circles in the cartesian $(x,y)$ plane
with arbitrary radius $R$ and center $(x_0,y_0)$:
\be
\label{qzm}
x(s)= x_0 + R \cos k s  \ , \qquad y(s) = y_0 + R  \sin k s
\ee
where $k\in \Zint$ and $s\in [0,2\pi]$ parameterizes the closed trajectory.
This becomes a solution of the equations of motion when
\be
\label{ons}
R = M\ y_0 / B\ ,\quad \tilde \rho = \frac{y_0\ k }{B}
\ee
We now truncate the path
integral over $x(s),y(s)$ to the ``mini-superspace'' of 
constant values of $(R,x_0,y_0)$. Using the natural
conformal-invariant integration measure, one obtains the 
mini-superspace approximation to the one-loop path integral,
\be
\label{imini}
I_{mini} = 
B^2 \int \frac{dx_0\ dy_0}{y_0^2}  
\int_{0}^{\infty} \frac{R\ dR}{y_0^2}
\int_{0}^{\infty}
\frac{d\tilde\rho}{\tilde \rho}\  
\exp(- S_E)
\ee
where $S$ is the classical Euclidean action of the configuration \eqref{qzm}
(assuming $0<R<y_0$)
\be
\label{s1}
S_E = \int_0^{2\pi} \ ds ~ {\cal L} =
\pi \left[ \frac{M^2\ \tilde\rho}{\sqrt{y_0^2-R^2}}
+  \frac{k^2\ R^2}{\tilde \rho \sqrt{y_0^2-R^2} }
+ 2\ B\ k \left( 1 - \frac{y_0}{\sqrt{y_0^2-R^2}} \right) \right]
\ee
By conformal invariance, $S$ depends only on the
ratio $R/y_0$ and $\tilde\rho/y_0$. The action \eqref{s1} is 
stationary with respect to the radius $R$ at 
\be
\langle R\rangle = \frac{1}{k} \sqrt{ M^2 \tilde\rho^2 - 2\  k\  B\
  y_0 
\ \tilde\rho + 2\  y_0^2\  k^2}
\ee
for any $\rho$; the resulting action at this saddle point action becomes
\be
\label{sR}
S_{\langle R\rangle} 
= \pi\left( 
2~k~B - \sqrt{\frac{k y_0~(2B \tilde\rho- k y_0)}{\tilde\rho^2} - M^2} \right)
\ee
Note in particular that in contrast to the flat space case,
the integral over the radius modulus $R$ never leads
to a pole in the $\tilde\rho$ Schwinger parameter plane.

If instead one first integrates over the Schwinger parameter  
$\tilde\rho$, one finds that $S$ is stationary at
 $\langle\tilde\rho\rangle=k R/M$, with saddle point action
\be
\label{sr}
S_{\langle\tilde\rho\rangle} =  \frac{2\pi k}{\sqrt{y_0^2-R^2}}
\left[ M~R + B \left( -y_0 + \sqrt{y_0^2-R^2} \right) \right]
\ee

In either case, extremizing \eqref{sr} with respect to
 $R$, or \eqref{sR} with respect to
$\tilde\rho$,  the resulting classical action becomes
\be
S_{\langle{R,\tilde\rho}\rangle} = 2\pi k \left( B - \sqrt{B^2-M^2} \right)
\ee
with saddle-point values
\be
\label{sad}
\langle R\rangle=M \frac{y_0}{B}, \qquad \tilde\rho=k \frac{y_0}{B}
\ee 
in agreement with the on-shell values \eqref{ons}.
In the one-instanton case $k=1$, this indeed reproduces the 
result \eqref{sc1} for $k=1$.

The advantage of this approach is that one can now evaluate the 
contribution of fluctuations (in the mini-superspace sector)
around the saddle point \eqref{sad}. Varying the action \eqref{s1}
with respect to both $(R,\rho)$, the one-loop determinant
around the Gaussian saddle point is given by
\be
\det = - \left[ \frac{4\pi}{y_0^2} \frac{B^3
 M}{B^2- M^2} \right]^2 < 0
\ee
The zero-modes arising from conformal invariance contribute a 
factor of volume. 
Assuming that higher fluctuation modes around the ``quasi-zero-mode 
trajectories'' \eqref{qzm} all have positive eigenvalues, one 
obtains a negative fluctuation determinant, implying that the
saddle point \eqref{sad} indeed contributes to the imaginary
part of the one-loop amplitude. Finally, including the summation
measure \eqref{imini}, one obtains, in the saddle point approximation,
\be
I_{mini} \sim i \ V \frac{B^2-M^2}{B} \sum_{k=1}^{\infty} \frac{1}{k} 
\exp\left[ - 2\pi k \left( E - \sqrt{E^2-M^2} \right) \right]
\ee
where $V=\int dx_0 dy_0/y_0^2$ is the regularized volume of $AdS_2$.
Identifying $B=E$ and taking the limit $B\gg M$, we recover precisely 
the first term of the exact field theory result \eqref{imli2}
 computed in the next Section
 (the other term in \eqref{imli2} can be seen to arise from
quantum corrections around the saddle point).

\section{One-loop vacuum amplitude in quantum field theory}
\label{oneloop}
Having reached a semi-classical understanding of the pair production
process, we now turn to a quantum field theoretical treatment, and evaluate 
the one-loop vacuum amplitude for a charged, spinless particle in
$AdS_2$, with particular emphasis on its imaginary part. 
Following the approach in \cite{Comtet:1984mm}, we start
from the Euclidean problem of a charged particle on the hyperbolic
plane $H_2$, and analytically continue to the Lorentzian problem
of interest\footnote{We thus deviate from \cite{Comtet:1984mm}, which
deals with a different analytic continuation leading to $dS_2$.}. 

\subsection{One-loop amplitude on $H_2$}
We have briefly recalled the semi-classics of the Landau problem
on the hyperbolic plane in Section \eqref{smsch} above. As observed in
\cite{Comtet:1986ki}, the semi-classical approximation to the energy
spectrum is in fact exact, up to a replacement of the energy $H$ by
$H+\frac14$. The spectrum of the magnetic Laplacian on $H_2$
\be
\label{maglap}
H_B = - y^2 \left[ \pa_y^2 + (\pa_x - i B/y)^2 \right] 
\ee
consists of a finite number of discrete Landau states, with energy
\begin{eqnarray}
\label{edisc}
H_n &=& 
%\frac{1}{ a^2} 
B^2 + \frac{1}{4} - (B-n-\frac{1}{2})^2 
\end{eqnarray}
where $0\leq n \leq B-\frac12$ is a non-negative integer, and a continuum of
delta-normalizable states with energy
\begin{eqnarray}
\label{econt}
H_\nu &=& %\frac{1}{ a^2} 
B^2 +\nu^2 + \frac14
\end{eqnarray}
for $\nu\in \Real^+$. Discrete states are present only for $B>1/2$.

Both the discrete states \eqref{edisc} 
and continuous states \eqref{econt} are infinitely degenerate with respect
to the quantum number $p_t$, which takes value in $\Real^+$ for
discrete states, and $\Real$ for continuous states (we assume $B>0$
throughout, and set the curvature radius to $a=1$). 
This infinite degeneracy can be regularized by
introducing an infrared volume cut-off $V$, leading to a degeneracy
\be
\rho_n = \frac{V}{2\pi} \left(B - n - \frac12 \right)
\ee
for the discrete states, and a
density of states 
\be
\rho_c(\nu) = \frac{V}{2\pi} 
\frac{\nu\ \sinh 2 \pi \nu }{ \cosh 2 \pi \nu + \cos 2 \pi B  }
\ee
in the continuum \cite{Comtet:1984mm}. This allows us to compute
the heat kernel for the magnetic Laplacian \eqref{maglap},
%The corresponding wave functions are recorded in \cite{Comtet:1984mm}.
\be
\label{kb}
K_B(\tau) := \Tr\ e^{- \tau H_B} 
= \sum_{n=0}^{B-1/2} \rho_n\  e^{-\tau H_n} 
+ \int_0^{\infty} d\nu \ \rho_c(\nu)\ e^{-\tau H_\nu}
\ee
Using parity considerations as well as the identity
\be
\label{psi1}
\Im \left[ \psi\left(\frac12 +i\nu-B\right)+
\psi\left(\frac12+i\nu+B\right) \right]
=\pi \frac{\sinh 2 \pi \nu }{ \cosh 2 \pi \nu + \cos 2 \pi B}
\ee
where $\psi(x) = d \log \Gamma(x) /dx$ is the Euler $\psi$ function,
the contribution of the continuum can be rewritten as an integral
\be
\label{kbc}
K_{B;c}(\tau) = \frac{V}{4\pi^2 i} \int_{-\infty}^{\infty} 
d \nu ~ \nu ~ \left[ \psi \left(\frac12 +i \nu-B \right)
+ \psi\left(\frac12+i \nu+B\right) \right] e^{-\tau E_\nu}.
\ee
on the complete real axis $\nu\in\Real$. Now, observe that 
the integrand has poles when the argument 
of $\psi(z)$ is a negative integer, i.e. at:
\be
\label{nun}
\nu = i \left( n + \frac12 \pm B \right)\ ,\quad n=0,1,2,\dots
\ee
For small $B<1/2$, all poles lie in the upper half plane (see Figure 3).
As $B$ is increased past $n+1/2$, the pole $i(n+\frac12-B)$ crosses
over into the lower half plane. It can be checked that the residue
at the pole
is precisely equal to the contribution of the discrete state of
energy $E_n$. The contributions
of the discrete and continuous spectrum may thus be combined
by shifting the integration contour 
in the lower half plane so that it passes below the poles
at $\nu = i (n + \frac12 - B)$ for any $n\geq 0$.
The complete  heat kernel $K_B(\tau)$ of the magnetic Laplacian 
may thus be concisely written as
\be
\label{KC}
K_B(\tau) = \frac{V}{4 \pi^2 i}  
\int_{C=- i (B-\frac12+\epsilon) + \Real} 
d \nu ~\nu ~ \left[ \psi\left( \frac12 +i \nu-B\right)
+ \psi\left(\frac12 +i \nu+B \right) \right] \ e^{-\tau E_\nu} .
\ee
The one-loop vacuum amplitude, or free energy, can now be
obtained from the heat kernel by integrating over Schwinger
time $\tau$,
\be
W_B = \Tr\ \log\left( H_B + M^2 \right) = 
- \int_0^{\infty} \frac{d\tau}{\tau} K_B(\tau) \ e^{-\tau M^2} 
\ee
(More precisely, this equation applies to the differences 
$W_B-W_0$ and $K_B - K_0$).

\FIGURE{
%\hspace*{3cm}
\epsfig{file=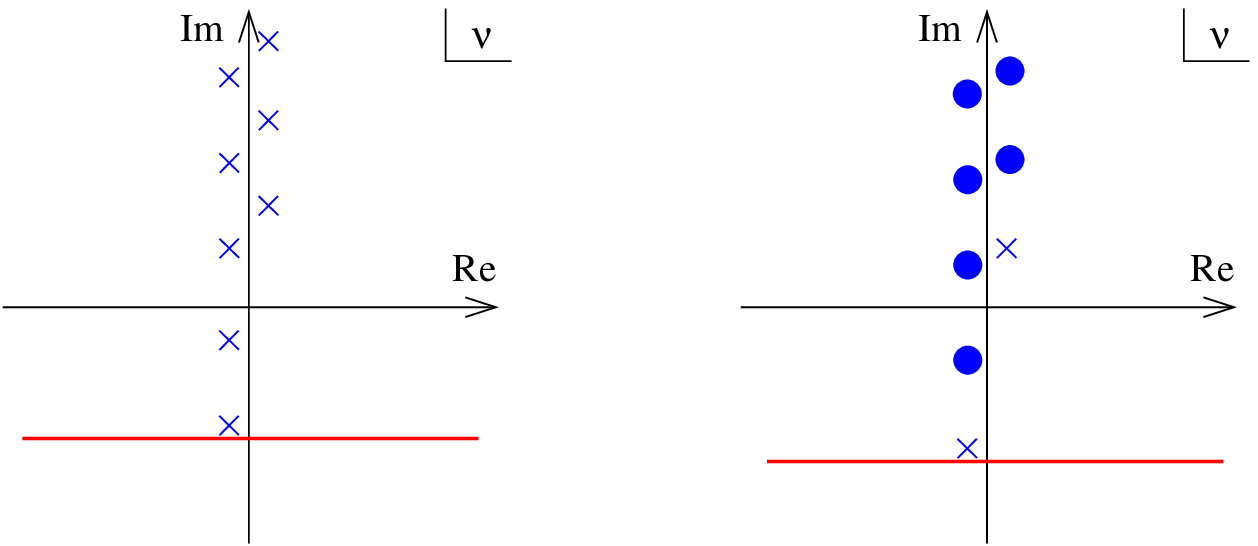,height=5cm}
%\hspace*{3cm}
\caption{The analytic structure of the heat kernel
for the magnetic Laplacian
in the $\nu$ plane. On the left, the poles and integration contour
for the scalar charged particle, on the right for the
spin 1/2 particle to be discussed in Section 3.4. 
The integration contour $C$ (in bold red)
sums the contribution of both
the discrete and continuous states. The crosses (in blue) denote
single poles, the (blue) dots denote double poles. The poles all lie
on the imaginary axis, but have been slightly separated for clarity.
}}

\subsection{Analytic continuation to $AdS_2$}
\label{ancont}
After first quantization, the world-line Hamiltonian \eqref{hpoin}
for an electron in (the Poincaré patch of) $AdS_2$ becomes
the Klein-Gordon equation $(H_E+M^2) \phi = 0$
on the complex wave function $\phi(y,t)$ of the electron, where
\be
\label{helec}
H_E := -y^2 \left[ \pa_y^2 - (\pa_t - i E/y)^2 \right]
\ee
is the Klein-Gordon operator for a charged field on $AdS_2$ 
with a constant electric field. Just as in the magnetic problem,
the one-loop amplitude $W_E = \log\det (H_E + M^2)$
can be obtained from the heat kernel by
integration with respect to Schwinger time,
\be
W_E = - \int_0^{\infty} \frac{d\tau}{\tau} K_E(\tau) e^{-\tau M^2} 
\ ,\quad
K_E(\tau) := \Tr\ e^{- \tau H_E} 
\ee
While we could in principle construct $K_E(\tau)$ from 
the density of continuous states in the potential $V(y)$ in \eqref{hpoinc}
(or from the density of continuous and discrete states in the potential
$V(\sigma)$ in \eqref{hglob}),
we choose instead to obtain it by analytic continution of the magnetic
heat kernel \eqref{KC}.

For this purpose, notice that the Schwinger Hamiltonian \eqref{helec}
is turned into the Landau Hamiltonian \eqref{maglap} by identifying
\be
x = -i t\ ,\quad B = i E \quad ,\quad H = - M^2
\ee
Under this analytic continuation, the discrete states \eqref{edisc}
become states with imaginary mass squared $M^2$, and are {\it not} 
part of the delta-normalizable spectrum of the Schwinger Hamiltonian 
\eqref{helec}. On the other hand, the continuous states \eqref{econt}
of the magnetic problem yield a continuum of states with mass squared
below the generalized Breitenlohner-Freedman bound 
$M^2 < E^2 - \frac14$, which are not delta-normalizable near $y=0$,
but are nevertheless part of the physical spectrum, and indeed
are responsible for the pair creation process. 

In order to obtain the heat kernel $K_E(z)$ for the electric problem,
let us start from the magnetic problem \eqref{KC}, set $B= e^{i\theta} E$
and increase $\theta$ from $0$ to $\pi/2$. For $\theta=0$, the analytic
structure of the integrand in the $\nu$ plane consists of poles \eqref{nun},
a finite number of which lie in the lower half plane.
As $\theta$ is increased
up to $\pi/2$, all the poles migrate to the sunny upper-half $\nu$ plane.
We can therefore lift the contour $C$ to the real axis, and obtain
\be
\label{KE1}
K_E(\tau) = \frac{V}{4 \pi^2 i}  
\int_{\Real} 
d \nu ~\nu ~ \left[ \psi\left( \frac12 +i (\nu-E) \right)
+ \psi\left(\frac12 +i (\nu+E) \right) \right] \ e^{-\tau E_\nu} .
\ee
Reversing the argument which took us from \eqref{kb} to \eqref{kbc},
this may be rewritten as an integral over the continuous spectrum
of the Klein-Gordon operator \eqref{helec}
\be
\label{KE2}
K_E(\tau) = 
\int_0^{\infty} 
d \nu \rho_E(\nu)  \ e^{-\tau E_\nu}
\ee
where the density of states\footnote{This applies to states with real $\nu$, 
i.e. below the BF bound $M^2<E^2-1/4$. } is now given by
\be
\label{trho}
\rho_E(\nu) = \frac{V}{2\pi} 
\frac{\nu\ \sinh 2 \pi \nu }{ \cosh 2 \pi \nu + \cosh 2 \pi E  }
\ee
The one-loop amplitude is thus given by the integral over the Schwinger
parameter $\tau$,
\be
W_E =  - \int_{0}^{\infty} \frac{d\tau}{\tau}
\int_{0}^{\infty} d\nu \ \rho_E(\nu) \ 
e^{- \tau (M^2-E^2+\nu^2+\frac14)}
\ee
For small electric field $E^2<M^2+1/4$, the one-loop amplitude
is well defined and real. For large electric field however,
the integral over $\tau$ has  an infrared divergence when
$\nu < \Delta$, where $\Delta=\sqrt{E^2-M^2-\frac14}$.
This divergence is analogous to the tachyonic divergence of
the bosonic string, and is expected to give an imaginary contribution 
to the one-loop amplitude. In order to extract the imaginary part,
it is convenient to expand the density of states as the difference
of geometric series
\be
\label{diffgeo}
\rho_E(\nu) =  \frac{V \nu}{2\pi}   
\sum_{k=1}^{\infty} (-1)^k\
\left[  e^{- 2 \pi (E+\nu) k} - \ e^{-2 \pi (E-\nu) k} \right]
\ee
which converge in the region of interest $0<\nu<\Delta<E$.
Using parity considerations once again, the imaginary part of the
one-loop amplitude may be written as
\be
\Im\left( W_E \right) =  - \frac{V}{2\pi}
\Im\left\{ 
\sum_{k=1}^\infty
(-1)^k
\int_{0}^{\infty} \frac{d\tau}{\tau}
\int_{-\Delta}^{\Delta} \nu \ d\nu 
\exp\left[-2\pi  (E+\nu) k + \tau (\Delta^2-\nu^2) \right] \right\}
\ee
The imaginary part is not affected by extending the integration
range to the full $\nu$ axis\footnote{A subtlety with this prescription
is that the sum over $k$ will only become convergent after performing
the integral over $\nu$.}. The $\nu$ integral now becomes Gaussian,
and yields 
\be
\Im\left(W_E\right) =  \frac{V}{2\pi}
\Im\left\{ \sum_{k=1}^\infty
(-1)^k\ k\ 
\int_{0}^{\infty} \frac{d\tau}{\tau}
\left(\frac{\pi}{\tau}\right)^{3/2}
\exp\left[ -2\pi k E + \frac{\pi^2 k^2}{\tau} + \tau \Delta^2 \right]
\right\}
\ee
The remaining integral over $\tau$ is of Bessel type, however it
diverges at both ends of the $\tau$ integration range. Generalizing
the procedure described in \cite{Marcus:1988vs} for regularizing 
divergent integrals,
we define it by analytic continuation of the integral
\be
\int_{0}^{\infty} \frac{d\tau}{\tau^{5/2}} \exp\left( - a\tau - b/\tau \right)
%&=& 2 \left(\frac{a}{b}\right)^{\frac34} K_{3/2} \left(
%2\sqrt{ab} \right) \\
=  \frac{\sqrt\pi}{2 b^{3/2}} \left( 1 + 2 \sqrt{ab} \right)
e^{-2 \sqrt{ab}}
\ee
which is convergent and real for $\Re(a)>0,\Re(b)>0$. Analytically 
continuing \footnote{An alternative prescription would be to rotate $a$ 
and $b$ oppositely, but the result would not have the correct
semi-classical limit.}
$(a,b) \to e^{i\pi} (a,b)$, we find a purely imaginary result,
\be
\int_{0}^{\infty} \frac{d\tau}{\tau^{5/2}} \exp\left( a\tau + b/\tau \right)
 =  \frac{i \sqrt\pi}{2 b^{3/2}} \left( 1 - 2 \sqrt{ab} \right)
e^{2 \sqrt{ab}}
\ee
Using this prescription, we obtain
\be
\Im\left(W_E\right) =  \frac{V}{4\pi^2} \sum_{k=1}^\infty
(-1)^k
\frac{1}{k^2} \left( 1- 2\pi k \Delta \right)
\exp\left[ - 2\pi k \left( E -\Delta \right) \right]
\ee
This provides our final result for the pair production
rate of spinless charged particles in $AdS_2$,
\be
\label{imli2}
\Im \left( W_E \right)=
\frac{V}{4\pi^2} \left[
\Li_2 \left( -e^{-2 \pi (E - \Delta )} \right)
- 2\pi \Delta\ 
\Li_1 \left(- e^{-2 \pi\left(E- \Delta \right)} \right)
\right]
\ee
\vskip2mm
\noindent where $\Li_n(x)=\sum_{k=1}^\infty x^k/k^n$ is the
standard poly-logarithm function, and we recall that 
$\Delta= \sqrt{E^2-M^2-\frac14} > 0$.
Several comments are in order:
\begin{itemize}
\item For large electric field $E\gg M$, the second term
dominates over the first. In this limit, the 
effect of curvature can be neglected. Approximating $\Delta \sim
E - M^2/(2E)$, we recover the standard Schwinger result in 
two flat dimensions,
\be
\Im W_E \to \frac{V}{2\pi} \sum_{k=1}^{\infty} \frac{E}{k}
\exp\left( - \pi k M^2 / E \right)
\ee
\item Semi-classically, the result \eqref{imli2} shows contributions
of saddle points with action $2\pi k (E - \Delta)$: this is precisely
the action of the closed periodic orbits of the magnetic problem 
which control the tunneling under the barrier of the electric problem.
Our mini-superspace analysis in Section 2.3 precisely reproduces the
logarithmic term in \eqref{imli2}, which dominates in the semi-classical
limit over the dilogarithm.

\item The di-logarithm however dominates near the pair production 
threshold $E^2 \sim M^2+\frac14$, and leads to a finite non-zero result. The
imaginary part of the amplitude is therefore discontinuous at the
threshold.
\end{itemize}

\subsection{Spin 1/2 case}
The case of charged spin 1/2 case can be treated by very similar techniques,
using the fact that the square of the Dirac operator is a Klein-Gordon type
operator on the tensor product $L_2(AdS_2) \otimes \mathbb{C}^2$. 
The spectrum in the Euclidean
magnetic case has been derived in \cite{Comtet:1984mm}, and consists of
a finite number of discrete Landau states, with energy
\begin{eqnarray}
%\label{edisc}
H_{n,1/2} &=& %\frac{1}{ a^2}
B^2 - (n - B)^2 
\end{eqnarray}
where $0\leq n < B$ is a non-negative integer, and a continuum of
delta-normalizable states with energy
\begin{eqnarray}
%\label{econt}
H_{\nu,1/2} &=& %\frac{1}{ a^2} 
B^2 +\nu^2 
\end{eqnarray}
for $\nu\in \Real^+$. All discrete states except for the ground state $n=0$
are doubly degenerate. In contrast to the spin 0 case, discrete states
exist for arbitrarity small $B$.

The density of states of the continuous series can be found
in \cite{Comtet:1984mm}
\be
\rho_{E,1/2}(\nu) = \frac{V \nu}{\pi} 
\frac{\sinh 2\pi \nu}{\cosh 2\pi \nu - \cosh 2\pi E}
%:=\tilde \rho_c(\nu) + \frac{V \nu}{\pi} 
\ee
and rewritten in the same way as in \eqref{psi1},
\be
\rho_{E,1/2}(\nu) = \frac{V \nu}{\pi^2} \ \Im\left[
\psi\left(i\nu-B\right)+\psi\left(i\nu-B+1\right)+
\psi\left(i\nu+B\right)+\psi\left(i\nu+B+1\right) \right]
\ee
Again, the contributions of the discrete states to the heat kernel
or zeta function can be absorbed by shifting the contour $C$ so that
it lies below all the poles in the lower $\nu$ plane. Under analytic
continuation to the electric case, all poles move into
the upper half plane as in the spinless case.
The imaginary part of the free energy can be computed by
expanding the density of states as a sum of two convergent geometric
series,
\be
\rho_{E,1/2}(\nu) = \frac{V\nu}{\pi}
\sum_{k=1}^{\infty}   \left[ e^{- 2 \pi (E+\nu) k} 
- \ e^{-2 \pi (E-\nu)} \right]
\ee
We state the final result,
\be
\label{imli2f}
\Im W_E =
\frac{V}{2\pi^2} \left[
\Li_2 \left( e^{-2 \pi (E - \Delta_{1/2} )} \right)
- 2\pi \Delta_{1/2}\ 
\Li_1 \left( e^{-2 \pi\left(E- \Delta_{1/2} \right)} \right)
\right]
\ee
where $\Delta_{1/2}=\sqrt{E^2-M^2}$.
As in flat space, we observe that the order $k$ instanton terms all
come with the same sign, in contrast with the bosonic (spinless) case.

\section{Summary and Discussion}

In this work, we have performed a detailed analysis of the quantum
fluctuations of charged particles in $AdS_2$, both at a semi-classical
and field-theoretic level, with a particular emphasis on the production 
of charged pairs by the Schwinger mechanism. Let us recapitulate the
main points of our analysis, with a few more remarks:

\begin{itemize}

\item Due to the confining property of the $AdS$ geometry, 
pair production only takes place when the electric field exceeds the 
threshold $E^2 > M^2 + 1/4$. This is in contrast to flat 
space \cite{Schwinger:1951nm}, where 
pair production occurs (albeit non-perturbatively in $E$) for 
arbitrarily small electric field, which consequently relaxes to zero.
This property is crucial for the existence of stable supersymmetric 
$AdS_2 \times S^2$ vacua, as occur near the horizon 
of extremal Reissner-Nordstr\"om black holes.

\item One way of rephrasing this property is that 
the standard Breitenlohner-Freedman bound $M^2 \ge -1/4$
for a neutral particle in $AdS_2$ \cite{Breitenlohner:1982jf} 
is raised in the presence of an electric field $E$ to 
\be
\label{bfe}
M^2 \geq -\frac14 + E^2 
\ee
This is simply the normalizability condition for the 
wave function of a charged particle in the Poincar\'e patch
near the boundary $y=0$ of $AdS_2$. Pair production takes place
at large electric field when this condition is no longer met,
and is qualitatively similar to tachyon condensation. This phenomenon will 
be generic for $AdS$-spaces with a flux proportional to the volume, in the
presence of branes coupling to the flux, although a simple field theory 
approach may no longer be applicable.

\item As in flat space, pair production takes place via tunneling 
in the potential governing the radial motion. It is thus governed
by the Euclidean trajectory of a charged particle on the hyperbolic
plane $H_2$ with a constant magnetic field. The production rate 
in the Poincar\'e patch is given semi-classically by
\be
\label{gs}
\Gamma \sim \exp\left[ - 2\pi \left( E - \sqrt{E^2-M^2-\frac14}
\right) \right]
\ee
and reduces to the flat space 
answer $\exp( - M^2 / (2E) )$ when the 
ratio $(M^2+(1/4))/E^2$ becomes small. 

\item The above observations are confirmed by a detailed analysis of the
one-loop amplitude $W_E$. In contrast to the flat space case, 
the imaginary part of $W_E$
does {\it not} receive contributions from poles 
of the heat kernel in the imaginary Schwinger time domain, but rather
from unstable saddle points. This can be seen in the mini-superspace
version of the one-loop integral, which shows that
the quadratic action around Euclidean periodic trajectories has no 
zero-modes, but has one unstable mode. 
%Instead, the density of
%states (or resolvent) has poles at imaginary energy, a finite number
%of which correspond to the Landau states for a magnetic field on the
%hyperbolic plane \cite{Comtet:1986ki}. 
%The saddle point contribution \eqref{gs} comes about
%after Poisson resummation on the integer $n$ labelling the pole in
%\eqref{nun}.

\item Our result for the imaginary part of the one-loop amplitude
assumes a specific choice of vacuum. 
Due to existence of time-like boundaries, it is also necessary
to specify boundary conditions for light particles
which can escape to the boundary or be injected from it.
In Appendix A, we propose a different regularization scheme which leads to 
a different imaginary part, with two saddle points interfering
destructively at threshold. It would be desirable to 
understand the vacuum and boundary conditions implicit in both cases
in detail.

\item We have shown that the semi-classical tunneling rates agree
in the Poincar\'e patch and in global $AdS_2$. While we expect that the
known equality of the Poincar\'e and global vacua for neutral particles
carries over to the charged case, it would be interesting
to confirm this by constructing the one-loop amplitude in global $AdS_2$. 

\item Our computation has neglected gravitational backreaction. The
latter has been discussed long ago using an Euclidean approach to quantum 
gravity  \cite{Brown:1988kg}. While their results are in agreement with
ours at leading order in Newton's constant $G$  (see  Appendix B), 
they indicate that the tunneling action is corrected by gravitational 
effects to
\be
\label{acgrav}
S =  2 \pi (E-\sqrt{E^2-M^2}) + \frac{\pi G}{2} E
\left(-2 E + \frac{2 E^2-M^2}{\sqrt{E^2-M^2}} \right)
\ee
(with $E>0$) at first order in $G$. 
In particular, they seem to indicate that the threshold for particle
creation is further raised to $|E|> M + G M^2/4$.

\item As already stated, pair production is not expected to take place
in a supersymmetric setting. Nevertheless, there may exist charged BPS states
which just saturate the generalized Breitenlohner-Freedman bound \eqref{bfe}.
This is in particular so when $AdS_2$ 
arises as the near horizon geometry of a large number of BPS branes: 
an extra brane probe of the same type as those which created the background
will saturate the BPS bound \eqref{bfe}. Such states may be abundantly
produced and will cause AdS to fragment in several 
baby universes \cite{Maldacena:1998uz}. While the semi-classical
production rate has been computed in \cite{Maldacena:1998uz}, it is
important to note that restricting to the leading term in the
semi-classical approximation is inappropriate at the threshold
for pair production, where the  quantum corrections 
due to the dilogarithms in \eqref{imli2} become dominant. It is
also important to specify the spectrum and boundary conditions,
as the results \eqref{imli2} and \eqref{imli3} lead to very different
physics at threshold.

\item We expect our results to be useful in understanding perturbative
(heterotic or type II) string theory on $AdS_2\times S^2$. As pointed 
out in \cite{Israel:2004vv}, the partition function of string theory
on the hyperbolic deformation $Sl(2)/U(1)_k$ has remained elusive,
as it requires an understanding of characters of $Sl(2)/U(1)_k$ in an 
hyperbolic basis. Using the results in the present work, appropriately
generalized to higher spins, it should be possible to take a constructive
approach to this problem and get at the string one-loop amplitude by
summing the one-loop vacuum amplitude of each state in the spectrum. 

\item Charged particles in $AdS_2$ also arise in the context
of open strings stretched between two D-branes in $AdS_3$\footnote{We
thank C. Bachas for pointing this out to us.}. Recall that
$AdS_3$ admits a class of D-branes with $AdS_2$ world-volume supporting
an electric field \cite{Bachas:2000fr}. 
Open strings stretched between two such D-branes
carry a net electric charge under the difference of the respective 
electric fields. The result for the annulus amplitude 
in the Euclidean case was computed 
in \cite{Lee:2001gh}, and bears a close resemblance with 
the heat kernel \eqref{kb} of the magnetic Laplacian on $H_2$. It would
be interesting to revisit the analytic continuation to the
Lorentzian case, and see whether the world-sheet duality
transformation used in getting this result may shed light on
the construction of a modular invariant partition function in 
the heterotic case.

\item Finally, it would be interesting to revisit the problem originally
envisaged in \cite{Comtet:1984mm} of the pair production of  charged particles
in two-dimensional de Sitter space. This amounts to reversing 
the interpretation of the space and time coordinates of 
the Poincar\'e patch of $AdS_2$,
and flipping the sign of $M^2$. The classical trajectories of a charged
particle in $dS_2$ are therefore obtained by reading Figure 1 for
tachyonic trajectories sideways. By analogy, we expect the upper
bound on the mass of particles in $dS_2$ to be lowered in the
presence of an electric field. It would be interesting to compute
the modifications to the cosmological particle production rate due
to the electric field\footnote{This has been 
studied very recently at a semi-classical
level including the gravitational field in \cite{Gomberoff}.}.
\end{itemize}

To conclude, we feel that a quantum analysis of electromagnetic phenomena 
in strongly curved space-times such as $AdS_2 \times S^2$ may be of 
great value for studying black hole physics or cosmology. 

\vspace{5mm}

\noindent {\it Acknowledgements:}
It is a pleasure to thank M. Berkooz, O. Domenico, D. Israel, 
C. Kounnas, M. Petropoulos and M. Rozali for useful discussions,
and especially C. Bachas for insightful 
comments on the first version of this
paper.

\appendix

\section{Zeta function regularisation}
In this appendix we discuss an alternative way to regularize
the one-loop amplitude, using zeta-function 
regularization \cite{Comtet:1984mm}.
This will turn out to yield a different imaginary part for 
the one-loop amplitude, which may be interpreted as arising
from a different choice of vacuum and boundary conditions. 

Indeed, one may obtain the free energy from the derivative 
at $z=0$ of the zeta function $\zeta_B(z)$
associated to the magnetic Laplacian,
\be
\label{wez}
W_B = - \zeta'_B(0)\ ,\quad \zeta_{B} (z) := \Tr\  (H_B + M^2)^{-z}
\ee
Using the same manipulations as in Section \ref{ancont}, 
we can rewrite $\zeta_B(z)$
as a contour integral along the same contour $C$ as in \eqref{KC},
\be
\label{ZB}
\zeta_B(z) = \frac{V}{4 \pi^2 i} \int_C \ d\nu  ~\nu ~ 
\left[ \psi\left( \frac12 +i \nu-B\right)
+ \psi\left(\frac12 +i \nu+B \right) \right]\ [w(\nu)]^{-z}
\ee
where we defined
\be
w(\nu) = \frac{1}{4}+B^2+\nu^2 + M^2 
\ee
As it stands however,  $\zeta_B(z)$ is only well-defined for
$\Re(z)>1$. In order to continue near $z=0$, one should substract 
the leading contribution at $\nu\to\infty$: approximating
$\psi(x)\sim \log x$, this is 
\be
\frac{V}{2 \pi^2 i}  \int_C d \nu\  \nu \log \nu
[w(\nu)]^{-z}
= \frac{V}{4 \pi } \frac{[w(0)]^{1-z}}{z-1}
\ee
Adding and subtracting this term, we may rewrite the heat kernel as
\bea
\label{ZBR}
\zeta_B(z) &=& 
 \frac{V}{4 \pi } \frac{[w(0)]^{1-z}}{z-1} \\
&&+  \frac{V}{4 \pi^2 i} \int_C \ d\nu  ~\nu ~ 
\left[ \psi\left( \frac12 +i \nu-B\right)
+ \psi\left(\frac12 +i \nu+B \right) - 2 \log \nu \right]\ [w(\nu)]^{-z}
\nonumber
\eea
which converges for $\Re(z)>-1/2$, and is therefore suitable 
to compute the free energy \eqref{wez}. 
%For what regards the
%heat kernel \eqref{kb}, the above subtraction corresponds to
%the contribution of the  flat space density of states 
%$\rho(\nu \to \infty) \sim V \nu / (2\pi)$.

\FIGURE{
\epsfig{file=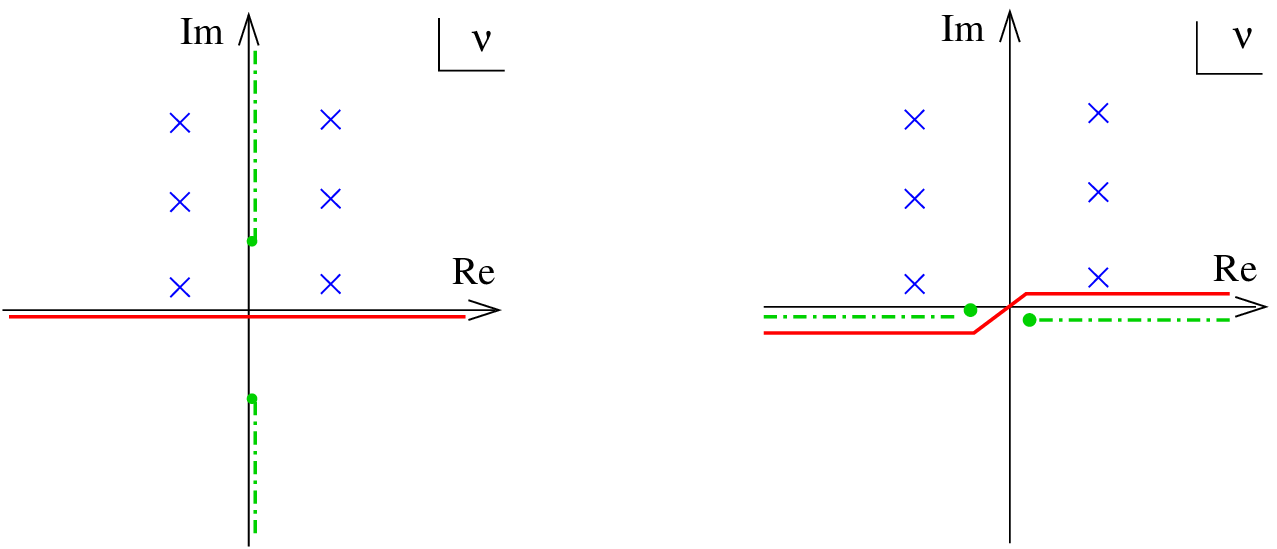,height=5cm} 
\caption{Analytic structure of the heat kernel and zeta function
and the electric Laplacian in the $\nu$ plane, for subcritical
(left) and supercritical (right) electric field. The green dotted line
denotes the cut from $\log w(\nu)$. 
}}

Let us now analytically continue the euclidean result by setting
$B= e^{i\theta} E$  and  increasing $\theta$ from $0$ to $\pi/2$.
as in Section \ref{ancont}. 
For $\theta=0$, in addition to the poles \eqref{nun},
there are two semi-infinite cuts 
along the imaginary axis, starting at $\nu= \pm i (
\sqrt{w(0)} + \Real^+ )$
in the lower and upper half plane respectively . As $\theta$ is increased
up to $\pi/2$, $w(\nu)$ rotates to
\be
\label{twnu}
\tilde{w} (\nu)= M^2  - E^2 + \frac14 +\nu^2
\ee
We now split the discussion according to the sign 
of $\tilde{w}(0)$ (see Figure 4):
\begin{itemize}
\item For  $M^2>E^2 -1/4$ (small electric field), the cuts 
stay on the imaginary axis at
$\nu = \pm i  ( \sqrt{M^2+1/4-E^2} + \Real^+)$.
We can therefore move 
the contour $C$ toward the real axis. Since $\tilde{w}(\nu)$
remains positive for any $\nu$,  we obtain
\be
\zeta_{E}(z) = \frac{V}{4 \pi} \frac{\tilde{w}(0)^{1-z}}{z-1}
+ \int_0^{\infty} d \nu  
\tilde \rho_E (\nu) \ [\tilde{w}(\nu)]^{-z}  
\ee
where $\rho_E(\nu)$ is the density of continuous states \eqref{trho}
in the electric problem,
and $\tilde \rho_E (\nu) = \rho_E (\nu) - \nu V /(2\pi)$.
The resulting free energy
\be
W_E = - \zeta'_E(0) = \frac{V}{4\pi} \tilde w(0) \left( 1 - \log  \tilde w(0)
\right) + \int_0^{\infty} d \nu  
\tilde \rho_E (\nu) \log \tilde{w}(\nu) 
\ee
is a real convergent integral, 
in agreement with the fact that there is no pair creation
for small electric field.

\item For $M^2+1/4<E^2$ (large electric field), the cuts at $\theta=\pi/2$
migrate to the real axis at $\nu = \pm   ( \sqrt{E^2-M^2-1/4} + \Real^+) $ 
when  $M^2<E^2-1/4$. The contour
$C$ now runs below the real axis for $\nu < E^2-M^2-1/4$, and
above the real axis for $\nu \geq E^2-M^2-1/4$. Noting that
$\tilde{w}(\nu)$ becomes negative for $\nu \in [  0, E^2-M^2-1/4 ]$,
we obtain
\begin{eqnarray}
\zeta_{E}(z) &=& \frac{V}{4 \pi} \frac{e^{i \pi (1-z)}
[-\tilde{w}(0)]^{1-z}}{z-1}
 + e^{-i \pi z} \int_0^{\sqrt{E^2-M^2-1/4}}
d \nu \tilde\rho_E(\nu) [-\tilde{w}(\nu)]^{-z} \nonumber \\
& & 
+ \int_{\sqrt{E^2-M^2-1/4}}^\infty
d \nu  \tilde\rho_E(\nu)  [w(z)]^{-z}.
\end{eqnarray}
where the (subtracted) density of states $\tilde\rho_E(\nu)$ is given by the 
same equation \eqref{trho} as before. Computing  $- \zeta'_E(0)$, we now
obtain both a real and imaginary part for the free energy,
\bea
\label{rew}
\Re(W_E) &=&  \frac{V}{4\pi} \tilde w(0) \left( 1 - \log  |\tilde w(0)|
\right) + \int_{\sqrt{E^2-M^2-1/4}}^{\infty} d \nu  \ 
\tilde \rho_E (\nu) \log \tilde{w}(\nu) \nn \\
&& \hfill + \int_0^{\sqrt{E^2-M^2-1/4}} d \nu \  
\tilde \rho_E (\nu) \log[- \tilde{w}(\nu)]  \\
\label{imw}
\Im(W_E) &=& \frac{V}{4} |\tilde{w}(0)| + \pi 
\int_0^{\sqrt{E^2-M^2-1/4}} d \nu \ \tilde \rho_E(\nu)
\eea
The volume term may be reabsorbed 
by restoring $\rho_E(\nu)$ into the integral.
\end{itemize}
As in the body of the paper, we may further process the imaginary
part of the amplitude \eqref{imw} by expanding the density of
states as a difference of geometric series \eqref{diffgeo}.
Integrating term by term, we obtain 
\bea
\label{imli3}
\Im \left( W_E \right) &=& - \frac{V \Delta}{4 \pi } 
\left[ \Li_1 \left(- e^{-2 \pi\left(E- \Delta \right)} \right)
+ \Li_1 \left( -e^{-2 \pi (E+ \Delta )} \right) \right]
\nonumber \\
& & +  \frac{V}{8\pi^2}
\left[ \Li_2 \left( -e^{-2 \pi (E - \Delta )} \right)
- \Li_2 \left(- e^{-2 \pi (E+ \Delta)} \right) \right] 
\eea
Similarly, for spin 1/2 fields, we obtain
\bea
\label{imli3f}
\Im \left( W_{E,1/2} \right) &=& - \frac{V \Delta_{1/2}}{2 \pi } 
\left[ \Li_1 \left( e^{-2 \pi\left(E- \Delta_{1/2} \right)} \right)
+ \Li_1 \left( e^{-2 \pi (E+ \Delta_{1/2} )} \right) \right]
\nonumber \\
& & +  \frac{V}{4\pi^2}
\left[ \Li_2 \left( e^{-2 \pi (E - \Delta_{1/2} )} \right)
- \Li_2 \left( e^{-2 \pi (E+ \Delta_{1/2})} \right) \right] 
\eea
In contrast to \eqref{imli2},\eqref{imli2f}, these results exhibit
the contribution of two saddle points with action $2\pi(E\pm\Delta)$.
Only the saddle point with action $2\pi(E-\Delta)$ has a semi-classical
interpretation as a tunneling process. In the framework of the
analysis of the one-loop amplitude in Section 3, the other
saddle point originates from negative Schwinger time. We believe
that the two results are equally valid, but for different choices
of vacua and boundary conditions. An important property of the
result \eqref{imli3},\eqref{imli3f} is that the imaginary part of the one-loop
amplitude vanishes at threshold. Applied to a supersymmetric context,
this indicates that BPS particles are not emitted in this vacuum.

\section{Gravitational back-reaction}
In this appendix, we briefly discuss the relation of our results for
pair creation in $AdS_2$ to the classic calculation of the
neutralization of the cosmological constant by brane 
creation \cite{Brown:1988kg} (see their
Appendix B for the two-dimensional case), 
to which we refer for many details.

Let us summarize the main results. In imaginary
time, pair creation of particles in $AdS_2$ amounts to the nucleation
of a bubble of AdS space of cosmological constant $\Lambda_i$ inside an
AdS space of cosmological constant $\Lambda_o$.  The Euclidean instanton
can be obtained by gluing two copies of $H_2$, in compatibility with
the Israel matching conditions. In the
semi-classical approximation, the pair production rate is proportional to
$e^{-S}$, where $S$ is the difference of the action of $AdS$ with (resp. 
without) the bubble,
\be
S =  2 \pi M \bar{\rho}
- \frac{4 \pi}{G} \log \frac{\bar{r}_i}{\bar{r}_o}
+ \frac{2 \pi \bar{\rho}}{G} (\Lambda_i \bar{r}_i-\Lambda_0 \bar{r}_0)
\ee
Here $M$ is the mass of the charged particle,
$\bar{\rho}$ is the classical proper radius of a particle
trajectory, and $G$ is Newton's constant.
The conformal factors of the metrics of the instanton
and the background are given in terms of inside and
outside radii $\bar r_{i}, \bar r_{o}$ as:
\begin{eqnarray}
\bar{r}_{i,o} &=& \frac{2}{\Lambda_{i,o} \bar{\rho}}
\left( 1-% \sigma_{i,o}
 \sqrt{1-\Lambda_{i,o} \bar{\rho}^2} \right) .
\end{eqnarray}
%For anti-de Sitter spaces, the only relevant instantons have
%$\sigma_{o,i}=+1$.
The equations of motion imply
the following equations for the cosmological
constants and the electric fields inside and outside
the particle trajectory
\begin{eqnarray}
\Lambda_{i,o} &=& \lambda + \frac{G}{2} E_{i,o}^2 \nonumber \\
E_o-E_i &=& % \epsilon 
e \nonumber \\
%K_o - K_i &=& \frac{k}{2} M \nonumber \\
- % \sigma_o 
(\bar{\rho}^{-2}-\Lambda_o)^{1/2}
+ % \sigma_i 
(\bar{\rho}^{-2}-\Lambda_i)^{1/2}
&=& \frac{G}{2} M
\nonumber \\
- %\sigma_o 
(\bar{\rho}^{-2}-\Lambda_o)^{1/2}
- %\sigma_i 
(\bar{\rho}^{-2}-\Lambda_i)^{1/2}
&=& -2 \frac{e}{2M} (E_i+E_o),
%\nonumber \\
%\sigma_{i,o} &=& sgn(\epsilon e E_o-e^2/2 \pm 1/4km^2)
\end{eqnarray}
where $e$ is the charge of the particle and $E_{i,o}$ are the
electric fields inside and outside the particle trajectories. 
Let us define the difference between the cosmological constants
as ($E>0$): 
\be
  \Lambda_i- \Lambda_o = \frac{G}{2} (E_i+E_o) (E_i-E_o)
 \equiv -G E.
\ee
Note that $E$ has to be positive, since
pair creation tends to reduce the cosmological constant.

In order to make contact with the semi-classical analysis in the bulk
of the paper, we will set $\Lambda_o = -1/a^2=-1$, and
expand the results in terms
of the gravitational coupling constant $G$. 
The quantity $E$ is roughly the background electric field
times the charge of the particle, and will coincide with our 
definition for the electric field $E$ in the bulk of the paper, in
the weak-gravity limit.

By extremizing the action with respect to the classical radius $\bar\rho$
we find:
\begin{eqnarray}
\bar{\rho} &=& (\Lambda_0+ 1/M^2(E+GM^2/4)^2)^{-1/2}.
\end{eqnarray}
To obtain the particle production rate, we  evaluate
the action for the instanton. Expanding in powers of the
gravitational coupling, we find at next-to-leading order
the result in Eq. \eqref{acgrav}.
In addition, we note that the value of our radius
variable $R$ at the extremum of the action is:
%\bar{\rho} %\frac{R/y_0}{\sqrt{1-(R/y_0)^2}}
\be
\frac{R}{y_0} = \frac{\bar{\rho}}{1+\bar{\rho}^2} 
= \frac{M}{|E|-G M^2/4},
\ee
While this agrees with the value \eqref{ons} found in the bulk of the paper 
at small $G$, it implies that the instanton exists only when
\begin{eqnarray}
|E|> M + G M^2/4,
\end{eqnarray}
The threshold for pair production is thus slightly raised by 
gravitational effects, as compared to the generalized 
Breitenlohner-Freedman bound \eqref{bfe}.
It would be interesting to extend this analysis to the case
of charged BPS particles.


\begin{thebibliography}{00}


%\cite{Gibbons:1982ih}
\bibitem{Gibbons:1982ih}
G.~W.~Gibbons,
``Antigravitating Black Hole Solitons With Scalar Hair In N=4 Supergravity,''
Nucl.\ Phys.\ B {\bf 207} (1982) 337;
%%CITATION = NUPHA,B207,337;%%

%\cite{Gibbons:1987ps}
%\bibitem{Gibbons:1987ps}
G.~W.~Gibbons and K.~i.~Maeda,
``Black Holes And Membranes In Higher Dimensional Theories With Dilaton
Fields,''
Nucl.\ Phys.\ B {\bf 298} (1988) 741;
%%CITATION = NUPHA,B298,741;%%

%\cite{Garfinkle:1990qj}
%\bibitem{Garfinkle:1990qj}
D.~Garfinkle, G.~T.~Horowitz and A.~Strominger,
``Charged Black Holes In String Theory,''
Phys.\ Rev.\ D {\bf 43}, 3140 (1991)
[Erratum-ibid.\ D {\bf 45}, 3888 (1992)].
%%CITATION = PHRVA,D43,3140;%%


%\cite{Ferrara:1995ih}
\bibitem{Ferrara:1995ih}
S.~Ferrara, R.~Kallosh and A.~Strominger,
``N=2 extremal black holes,''
Phys.\ Rev.\ D {\bf 52}, 5412 (1995)
[arXiv:hep-th/9508072].
%%CITATION = HEP-TH 9508072;%%

%\cite{Moore:2004fg}
\bibitem{Moore:2004fg}
G.~W.~Moore,
``Les Houches lectures on strings and arithmetic,''
arXiv:hep-th/0401049.
%%CITATION = HEP-TH 0401049;%%


%\cite{Maldacena:1998uz}
\bibitem{Maldacena:1998uz}
J.~M.~Maldacena, J.~Michelson and A.~Strominger,
``Anti-de Sitter fragmentation,''
JHEP {\bf 9902} (1999) 011
[arXiv:hep-th/9812073].
%%CITATION = HEP-TH 9812073;%%


%\cite{Spradlin:1999bn}
\bibitem{Spradlin:1999bn}
M.~Spradlin and A.~Strominger,
``Vacuum states for AdS(2) black holes,''
JHEP {\bf 9911}, 021 (1999)
[arXiv:hep-th/9904143].
%%CITATION = HEP-TH 9904143;%%

%\cite{Kim:1998wy}
\bibitem{Kim:1998wy}
W.~T.~Kim,
``AdS(2) and quantum stability in the CGHS model,''
Phys.\ Rev.\ D {\bf 60}, 024011 (1999)
[arXiv:hep-th/9810055].
%%CITATION = HEP-TH 9810055;%%


%\cite{Gaiotto:2004pc}
\bibitem{Gaiotto:2004pc}
D.~Gaiotto, A.~Simons, A.~Strominger and X.~Yin,
``D0-branes in black hole attractors,''
arXiv:hep-th/0412179;
%%CITATION = HEP-TH 0412179;%%
%\cite{Gaiotto:2004ij}
%\bibitem{Gaiotto:2004ij}
D.~Gaiotto, A.~Strominger and X.~Yin,
``Superconformal black hole quantum mechanics,''
arXiv:hep-th/0412322.
%%CITATION = HEP-TH 0412322;%%

%\cite{Berkovits:1999zq}
\bibitem{Berkovits:1999zq}
N.~Berkovits, M.~Bershadsky, T.~Hauer, S.~Zhukov and B.~Zwiebach,
``Superstring theory on AdS(2) x S(2) as a coset supermanifold,''
Nucl.\ Phys.\ B {\bf 567}, 61 (2000)
[arXiv:hep-th/9907200].
%%CITATION = HEP-TH 9907200;%%




%\cite{Giddings:1993wn}
\bibitem{Giddings:1993wn}
S.~B.~Giddings, J.~Polchinski and A.~Strominger,
``Four-dimensional black holes in string theory,''
Phys.\ Rev.\ D {\bf 48}, 5784 (1993)
[arXiv:hep-th/9305083];
%%CITATION = HEP-TH 9305083;%%
%\cite{Lowe:1994gt}
%\bibitem{Lowe:1994gt}
D.~A.~Lowe and A.~Strominger,
``Exact four-dimensional dyonic black holes and Bertotti-Robinson space-times
in string theory,''
Phys.\ Rev.\ Lett.\  {\bf 73}, 1468 (1994)
[arXiv:hep-th/9403186].
%%CITATION = HEP-TH 9403186;%%





%\cite{Johnson:1994kv}
\bibitem{Johnson:1994kv}
C.~V.~Johnson,
``Heterotic Coset Models,''
Mod.\ Phys.\ Lett.\ A {\bf 10}, 549 (1995)
[arXiv:hep-th/9409062];
%%CITATION = HEP-TH 9409062;%%
%\cite{Berglund:1995dv}
%\bibitem{Berglund:1995dv}
P.~Berglund, C.~V.~Johnson, S.~Kachru and P.~Zaugg,
``Heterotic Coset Models and (0,2) String Vacua,''
Nucl.\ Phys.\ B {\bf 460}, 252 (1996)
[arXiv:hep-th/9509170].
%%CITATION = HEP-TH 9509170;%%

%\cite{Israel:2004vv}
\bibitem{Israel:2004vv}
D.~Israel, C.~Kounnas, D.~Orlando and P.~M.~Petropoulos,
``Electric / magnetic deformations of S**3 and AdS(3), and geometric cosets,''
arXiv:hep-th/0405213.
%%CITATION = HEP-TH 0405213;%%


%\cite{Strominger:1998yg}
\bibitem{Strominger:1998yg}
A.~Strominger,
``AdS(2) quantum gravity and string theory,''
JHEP {\bf 9901}, 007 (1999)
[arXiv:hep-th/9809027].
%%CITATION = HEP-TH 9809027;%%

%\cite{Gibbons:1998fa}
\bibitem{Gibbons:1998fa}
G.~W.~Gibbons and P.~K.~Townsend,
``Black holes and Calogero models,''
Phys.\ Lett.\ B {\bf 454}, 187 (1999)
[arXiv:hep-th/9812034].
%%CITATION = HEP-TH 9812034;%%



%\cite{Thompson:2003fz}
\bibitem{Thompson:2003fz}
D.~M.~Thompson,
``AdS solutions of 2D type 0A,''
arXiv:hep-th/0312156.
%%CITATION = HEP-TH 0312156;%%


%\cite{Strominger:2003tm}
\bibitem{Strominger:2003tm}
A.~Strominger,
``A matrix model for AdS(2),''
JHEP {\bf 0403}, 066 (2004)
[arXiv:hep-th/0312194].
%%CITATION = HEP-TH 0312194;%%

%\cite{Verlinde:2004gt}
\bibitem{Verlinde:2004gt}
H.~Verlinde,
``Superstrings on AdS(2) and superconformal matrix quantum mechanics,''
arXiv:hep-th/0403024.
%%CITATION = HEP-TH 0403024;%%

%\cite{D'Hoker:1983is}
\bibitem{D'Hoker:1983is}
E.~D'Hoker, D.~Z.~Freedman and R.~Jackiw,
``SO(2,1) Invariant Quantization Of The Liouville Theory,''
Phys.\ Rev.\ D {\bf 28}, 2583 (1983).
%%CITATION = PHRVA,D28,2583;%%

%\cite{Brill:1991rw}
\bibitem{Brill:1991rw}
D.~Brill,
``Splitting of an extremal Reissner-Nordstrom throat via quantum tunneling,''
Phys.\ Rev.\ D {\bf 46}, 1560 (1992)
[arXiv:hep-th/9202037].
%%CITATION = HEP-TH 9202037;%%

%\cite{Britto-Pacumio:1999ax}
\bibitem{Britto-Pacumio:1999ax}
R.~Britto-Pacumio, J.~Michelson, A.~Strominger and A.~Volovich,
``Lectures on superconformal quantum mechanics and multi-black hole  moduli
%spaces,''
arXiv:hep-th/9911066.
%%CITATION = HEP-TH 9911066;%%

%\cite{Comtet:1984mm}
\bibitem{Comtet:1984mm}
A.~Comtet and P.~J.~Houston,
``Effective Action On The Hyperbolic Plane In A Constant External Field,''
J.\ Math.\ Phys.\  {\bf 26}, 185 (1985).
%%CITATION = JMAPA,26,185;%%




%\cite{Nakatsu:1998st}
\bibitem{Nakatsu:1998st}
T.~Nakatsu and N.~Yokoi,
`Comments on Hamiltonian formalism of AdS/CFT correspondence,''
Mod.\ Phys.\ Lett.\ A {\bf 14}, 147 (1999)
[arXiv:hep-th/9812047].
%%CITATION = HEP-TH 9812047;%%

%\cite{Brezin:1970xf}
\bibitem{Brezin:1970xf}
E.~Brezin and C.~Itzykson,
``Pair Production In Vacuum By An Alternating Field,''
Phys.\ Rev.\ D {\bf 2} (1970) 1191.
%%CITATION = PHRVA,D2,1191;%%


%\cite{Pioline:2003bs}
\bibitem{Pioline:2003bs}
M.~Berkooz, and B.~Pioline,
``Strings in an electric field, and the Milne universe,''
JCAP {\bf 0311} (2003) 007
[arXiv:hep-th/0307280].
%%CITATION = HEP-TH 0307280;%%


%\cite{Comtet:1986ki}
\bibitem{Comtet:1986ki}
A.~Comtet,
``On The Landau Levels On The Hyperbolic Plane,''
Annals Phys.\  {\bf 173} (1987) 185.
%%CITATION = APNYA,173,185;%%



%\cite{Grosche:1989kj}\cite{Antoine:1990ew}
\bibitem{Grosche:1989kj}
C.~Grosche,
``Path Integration On The Hyperbolic Plane With A Magnetic Field,''
Annals Phys.\  {\bf 201} (1990) 258.
%%CITATION = APNYA,201,258;%%

%\cite{Antoine:1990ew}
\bibitem{Antoine:1990ew}
M.~Antoine, A.~Comtet and S.~Ouvry,
``Scattering On An Hyperbolic Torus In A Constant Magnetic Field,''
J.\ Phys.\ A {\bf 23} (1990) 3699.
%%CITATION = JPAGB,A23,3699;%%


%\cite{Danielsson:1998wt}
\bibitem{Danielsson:1998wt}
U.~H.~Danielsson, E.~Keski-Vakkuri and M.~Kruczenski,
``Vacua, propagators, and holographic probes in AdS/CFT,''
JHEP {\bf 9901} (1999) 002
[arXiv:hep-th/9812007].
%%CITATION = HEP-TH 9812007;%%


%\cite{Affleck:1981bm}
\bibitem{Affleck:1981bm}
I.~K.~Affleck, O.~Alvarez and N.~S.~Manton,
``Pair Production At Strong Coupling In Weak External Fields,''
Nucl.\ Phys.\ B {\bf 197}, 509 (1982);
%%CITATION = NUPHA,B197,509;%%
%\cite{Affleck:1981ag}
%\bibitem{Affleck:1981ag}
I.~K.~Affleck and N.~S.~Manton,
``Monopole Pair Production In A Magnetic Field,''
Nucl.\ Phys.\ B {\bf 194}, 38 (1982).
%%CITATION = NUPHA,B194,38;%%



%\cite{Berkooz:2004re}
\bibitem{Berkooz:2004re}
M.~Berkooz, B.~Pioline and M.~Rozali,
``Closed strings in Misner space: 
Cosmological production of winding strings,''
JCAP {\bf 0408}, 004 (2004)
[arXiv:hep-th/0405126].
%%CITATION = HEP-TH 0405126;%%




%\cite{Breitenlohner:1982jf}
\bibitem{Breitenlohner:1982jf}
P.~Breitenlohner and D.~Z.~Freedman,
``Stability In Gauged Extended Supergravity,''
Annals Phys.\  {\bf 144} (1982) 249.
%%CITATION = APNYA,144,249;%%

%\cite{Schwinger:1951nm}
\bibitem{Schwinger:1951nm}
J.~S.~Schwinger,
``On Gauge Invariance And Vacuum Polarization,''
Phys.\ Rev.\  {\bf 82} (1951) 664.
%%CITATION = PHRVA,82,664;%%

%\cite{Marcus:1988vs}
\bibitem{Marcus:1988vs}
N.~Marcus,
``Unitarity And Regularized Divergences In String Amplitudes,''
Phys.\ Lett.\ B {\bf 219}, 265 (1989).
%%CITATION = PHLTA,B219,265;%%

%\cite{Bachas:2000fr}
\bibitem{Bachas:2000fr}
C.~Bachas and M.~Petropoulos,
``Anti-de-Sitter D-branes,''
JHEP {\bf 0102}, 025 (2001)
[arXiv:hep-th/0012234].
%%CITATION = HEP-TH 0012234;%%


%\cite{Lee:2001gh}
\bibitem{Lee:2001gh}
P.~Lee, H.~Ooguri and J.~w.~Park,
``Boundary states for AdS(2) branes in AdS(3),''
Nucl.\ Phys.\ B {\bf 632} (2002) 283
[arXiv:hep-th/0112188];
%%CITATION = HEP-TH 0112188;%%
%\cite{Ponsot:2001gt}
%\bibitem{Ponsot:2001gt}
B.~Ponsot, V.~Schomerus and J.~Teschner,
``Branes in the Euclidean AdS(3),''
JHEP {\bf 0202}, 016 (2002)
[arXiv:hep-th/0112198];
%%CITATION = HEP-TH 0112198;%%
%\cite{Ribault:2002ti}
%\bibitem{Ribault:2002ti}
S.~Ribault,
``Two AdS(2) branes in the Euclidean AdS(3),''
JHEP {\bf 0305}, 003 (2003)
[arXiv:hep-th/0210248].
%%CITATION = HEP-TH 0210248;%%


%\cite{Brown:1988kg}
\bibitem{Brown:1988kg}
J.~D.~Brown and C.~Teitelboim,
``Neutralization Of The Cosmological Constant By Membrane Creation,''
Nucl.\ Phys.\ B {\bf 297} (1988) 787.
%%CITATION = NUPHA,B297,787;%%

%\cite{Teitelboim:1983ux}
\bibitem{Teitelboim:1983ux}
C.~Teitelboim,
``Gravitation And Hamiltonian Structure In Two Space-Time Dimensions,''
Phys.\ Lett.\ B {\bf 126}, 41 (1983).
%%CITATION = PHLTA,B126,41;%%

\bibitem{Gomberoff}
A.~Gomberoff, M.~Henneaux and C.~Teitelboim,
``Decay of the Cosmological Constant. Equivalence of Quantum Tunneling and Thermal Activation in Two Spacetime Dimensions'',
arXiv:hep-th/0501152.
%%CITATION = HEP-TH 0501152;%%



\end{thebibliography}
\end{document}